# In Vitro Durability - Pivot bearing with Diamond Like Carbon for Ventricular Assist Devices

Rosa Corrêa Leoncio de Sá [1,2]

Vladimir Jesus Trava Airoldi [3]

Tarcísio Fernandes Leão [1,2]

Evandro Drigo da Silva [1]

Jeison Willian Gomes da Fonseca [1]

Bruno Utiyama da Silva [1]

Edir Branzoni Leal [1]

João Roberto Moro [2]

Aron José Pazin de Andrade [1]

Eduardo Guy Perpétuo Bock [2]

[1] Center of Engineering in Assistance Circulatory (CEAC) of Institute Dante Pazzanese of Cardiology (IDPC). São Paulo / SP, Brazil;

[2] Department of Engineering Mechanical of Institute Federal of Education, Science and Technology of São Paulo (IFSP). São Paulo / SP, Brazil;

[3] Laboratory Associated of Sensors and Materials (LAS) of the Institute National for Space Research (INPE). São José dos Campos / SP, Brazil.

rosacldesa@gmail.com

# IOP | Biomedical Materials

**Abstract**

Institute Dante Pazzanese of Cardiology (IDPC) develops Ventricular Assist Devices (VAD) that can stabilize the hemodynamics of patients with severe heart failure before, during and/or after the medical practice; can be temporary or permanent. The ADV's centrifugal basically consist of a rotor suspended for system pivoting bearing; the PIVOT is the axis with movement of rotational and the bearing is the bearing surface. As a whole system of an implantable VAD should be made of long-life biomaterial so that there is no degradation or deformation during application time; surface modification techniques have been widely studied and implemented to improve properties such as biocompatibility and durability of applicable materials. The Chemical Vapour Deposition technique allows substrates having melting point higher than 300 °C to be coated, encapsulated, with a diamond like carbon film (DLC); The amorphous carbon film coated material provides the surface properties such as hardness, and improved self-lubricating and biocompatibility for specific application. IDPC together with The National Institute for Space Research coated pivots of the support system of ADV's implantable made of Poly-Ether-Ether-Ketone (PEEK) and titanium medical degree with DLC; since such a system operates continuously and in a hostile environment, i.e., in physical and chemical stress. Two more pivots were made to the study, one in pure medical titanium and other alumina ceramic; a bearing surface, the bearing was made of PEEK for each pivot. Then, there was the physical characterization of each component by electron microscopy of the surface and volume estimation scan on a scale of digital precision, before and after in vitro test durability. The test simulated the actual conditions in which the system of support remains while applying a ADV. The results have been prepared on a comparative basis; where you select the pivoting assembly which showed less deformation by abrasive wear.

rosacldesa@gmail.com

# IOP | Biomedical Materials

**Introduction**

Cardiovascular diseases are the leading cause of death in Brazil and in the world, accounting for 30 percent (%) of the deaths recorded in 2012 in the country. Today, biotechnology national the Dante Pazzanese Institute of Cardiology (IDPC) of São Paulo develops Assistive Devices Uni / Biventricular (ADV) in specific biomaterials for the application, which can provide consistent hemodynamic stability to a body with Heart Failure (IC) chronic, until it recovers [Andrade, 2012].

Prototypes in centrifugal ADV are basically consist of external structure with two cannulas access, an internal impeller suspended for pivotal bearing system and the motor housing in figure 1; the entry cannula is designed to be introduced directly into the left ventricle [Utiyama, 2012].

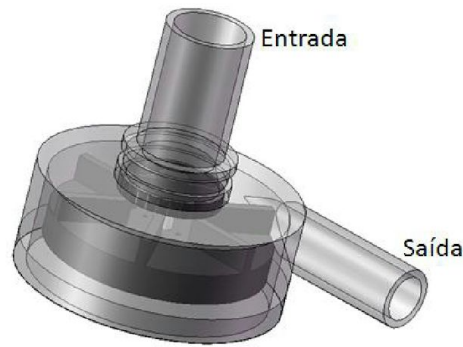

Figure 1. Centrifugal Pump apical aortic [adapted from Utiyama, 2012].

The components are made in biomaterials such as medical grade titanium; with the exception of the pivot bearing that must be made in biomaterials high durability for testing and interaction of applications with blood. Bock in 2007, studied the wear of various combinations of biocompatible materials of pivots and bearings for application to centrifugal blood pumps; in this study, he concluded that the bearing combination of in ceramic alumina with pivots polymer of ultra-high molecular weight polyethylene shows less abrasive wear during the actuation of prototype, where the rotor reaches about 3000 rpm to generate flow of 5 L / min to 100 mmHg [Bock, 2011].

Because of the need to increase reliability, biocompatibility and useful life of implantable medical devices, in particular ADV's; the pivot bearing system and the relevant biomaterials are being studied again to support a minimum application time of 10 years without showing action of deterioration or deformation.

rosacldesa@gmail.com

A bearing consists of the inner part, the pivot usually cylindrical and rotary motion and the fixed bearing surface, the bearing. Smooth surfaces are ideal for such applications since if irregularities certain spots of the surface may be closer to one another causing the lubricant does not play a part in friction conditions between the surfaces and the fluid; excessive friction between the surfaces leads to abrasion of the material with lower hardness and consequently the lower life of the system; in ADV's, this condition will lead to excessive breakdown of red blood cells, hemoglobin release in the blood plasma; ie high rate of hemolysis [Sa, 2015].

The prediction of the abrasive wear rate is controlled by plastic deformation rate and is presented from the Archard equation originally developed for the sliding wear.

Equation 1

$$W = K\frac{L}{H}$$

At where,
**W** is the wear volume loss rate = distance [$mm^3$/m];
**L** is the force applied [N] and;
**K** is the wear coefficient.

The suitability of equation 1 for abrasive wear due to Rabinowiczes; where the worn abrasive penetrates the surface at a depth that is proportional to the ratio between the hardness and applied force. The formed groove volume is completely removed and released into the circuit; There was thus forming the angle of attack as the geometry of the abrasive [Pintaúde 2002].

The technique Chemical Vapour Deposition (CVD) has been applied to modify surfaces of certain materials to improve their properties depending on the condition and the time of application. The coating surfaces with diamond like carbon film (DLC) by CVD is used increasingly in medical and spatial areas to customize materials subjected to physical stress, chemical, thermal or radiation for a long period. The DLC film is a metastable form of amorphous carbon containing a significant fraction of $sp^3$ type bonds; its properties suggest high mechanical hardness, chemical stability, high resistance to wear and corrosion and it is also biocompatible; pure DLC also has a bactericidal activity of 30% against Escherichia coli. [Trava-Airoldi, 2007].

rosacldesa@gmail.com

# IOP | Biomedical Materials

Here we present the result of a collaborative effort between the Center Engineering in Assistance Circulatory (CEAC) of IDPC and the Laboratory Associated of Sensors and Materials (LAS) of Institute National for Space Research (INPE). The components of the pivoting support system of ADV's centrifugal in different materials, uncoated and coated with DLC, have been tested In Vitro durability to select, in a comparative way, the pivoting assembly that have lower wear rate.

**Methods**

The making of pivots and bearings of the constituents of the pivoting support system was held at the Engineering Center for Circulatory Assistance through conventional machining technique. It was used polyether ether ketone (PEEK) to fabricate four support surfaces, bearings, for each pivotal in study; pure titanium grade medical for 2 pivots, PEEK for 1 pivot. The technique Plasma Enhanced Chemical Vapor Deposition (PECVD) It was applied to coat 2 pivots with DLC, 1 in titanium and the other in PEEK at INPE, see Figure 2.

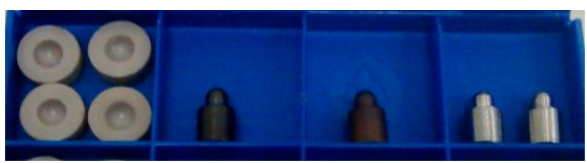

Figure 2. Bearings in PEEK left; followed by pivot in titanium with DLC , pivot in PEEK with DLC and pivot in pure titanium.

The DLC film was produced from methane ($CH_4$) and DC pulsed technique PECVD consisting of a discharge in a low pressure plasma using a switching power supply pulse for the generation of plasma and deposition of DLC films on substrates [Trava-Airoldi, 2007].

The pivot in alumina used in the studies per Bock, was also adopted. This study consisted of 4 tests In Vitro of durability carried out identically; Table 1 shows the sequence of the tests and the pairs of adopted materials.

Table 1. Pivot bearing sets ordered for testing in vitro durability.

| TESTS | PIVOT | | BEARING |
|---|---|---|---|
| 1 | Titanium with DLC | X | PEEK |
| 2 | PEEK with DLC | X | PEEK |
| 3 | Titanium medical | X | PEEK |
| 4 | Alumina | X | PEEK |

rosacldesa@gmail.com

The simulated test for 6 hours uninterrupted actual circumstances in which the support system remains during application on a pivoting ADV to select the set that has less wear or deformation. The test bench for the In Vitro durability test was composed of a digital scale, a milling machine (LFG, X63250) a PEEK bearing, a pivot, a support for the pivot and a lubricating solution (1/3 glycerin + 1 / 3 ethanol 70% + water) to simulate the density and viscosity of blood [Bock, 2011]. The wear rate, ie the volumetric loss on distance, or milligrams per meter [mg/m] of the material with lower hardness was determined by comparative statistical methods applying results of macroscopic and microscopic dimensional analysis and by estimating volume of the bodies before and after the tests. The density of the bearings and the pivots was recorded in milligrams for a digital precision scale (Adventurer, OHAUS) in IDPC; the micrograph of the surface was obtained by an Electron Microscope MIRA3- EGF Scan (TESCAN) at INPE.

**Results**

The presence of carbon nanocrystals with typical morphology of DLC nucleation "cauliflower" on the surface of the PEEK polymer can be confirmed by micrographic analysis FEG MIRA3 to 5 um resolution, in Figure 3.

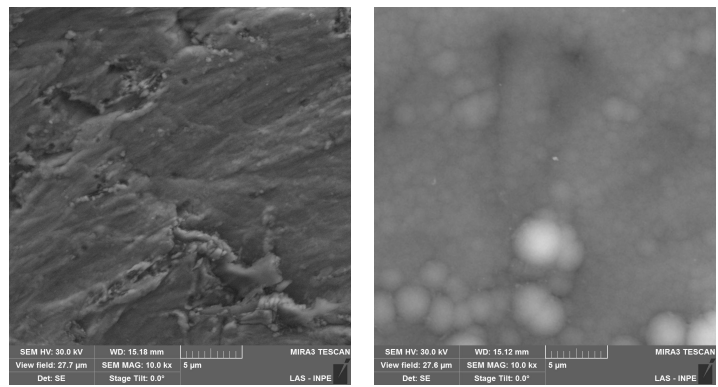

Figure 3. Micrographs obtained by FEG MIRA 3 of surface of Poly-Ether-Ether-Ketone before the after DLC coating.

Table 2 shows the weight values estimated by a precision balance, in milligrams, of each component under study; analyzes were performed before and after the bodies were subjected to in vitro assay.

rosacldesa@gmail.com

Table 2. Volume of mass of bodies in studies.

| ENSEMBLE | Component | Material | Mass early [mg] | Mass Final [mg] | Massa Equivalent [mg] |
|---|---|---|---|---|---|
| 1 | Bearing | PEEK | 186,7 | 186,3 | 0,4 |
| 1 | Pivot | Titânio Puro | 523,2 | 523 | 0,2 |
| 2 | Bearing | PEEK | 166,3 | 166,1 | 0,2 |
| 2 | Pivot | Titânio c/ DLC | 915,2 | 915,1 | 0,1 |
| 3 | Bearing | PEEK | 182,2 | 182,0 | 0,2 |
| 3 | Pivot | PEEK c/ DLC | 145,5 | 145,4 | 0,1 |
| 4 | Bearing | PEEK | 189,6 | 189,2 | 0,4 |
| 4 | Pivot | Alumina | 178,84 | 178,81 | 0,03 |

The micrographs of the surface under study revealed the impressive remains and traces resulting from abrasive wear; Figure 4 displays some aspects studied the PEEK bearing surface; before being subjected to wear and after in vitro assay.

**BEFORE *IN VITRO***     **AFTER *IN VITRO***

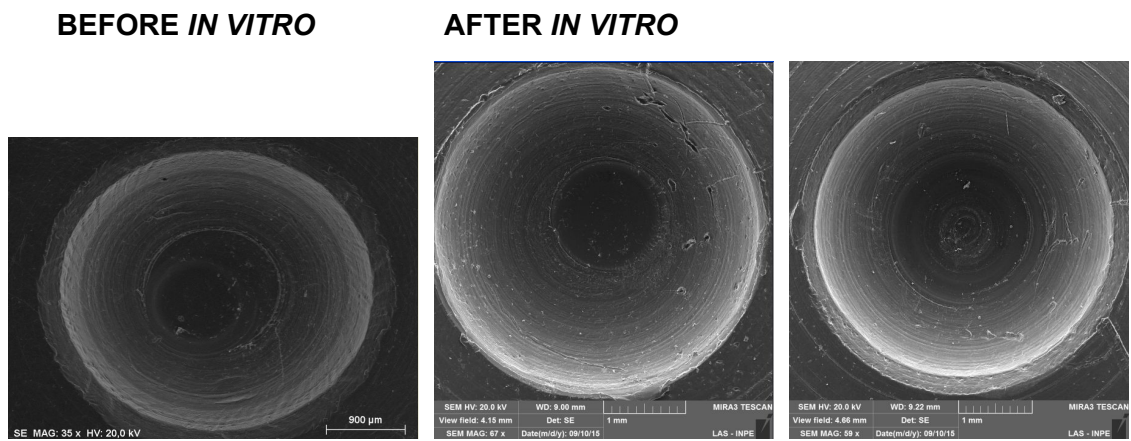

Figure 4. Bearing in virgin PEEK. Left, micrograph obtained before the test; followed by Group 2 constituents bearings (pivot titanium DLC) and 4 (pivot alumina) after the test.

**Discussion**

The PECVD technology proposed in this work was FOR maximize the durability of the pivotal system failure prevention by BSCI on the surface by abrading. The well adhered DLC coating tends to make the pivots on the surface smoother pivotal system with greater

**IOP | Biomedical Materials**

hardness and titanium; such properties are very promising for plain bearings, since if there are irregularities on the surface of a pivoting system, certain points may be closer to one another and the lubricant, in this case, blood can not play its role in conditions friction. Consequently, the biofuncionabilidade the device is compromised and the Normalized Hemolysis index will be high and unsuitable for ADV's.

**Conclusion**

The set 4, PEEK bearing and pivot alumina was the group that presented lower mass loss of the abrasive component, 0.03 mg; therefore, future studies should be made bearing finish PEEK with DLC to make a new durability test in order to reduce the mass loss of the bearing; as in the tests in conjunction with the pivot coated with DLC, the PEEK bearing showed a lower weight loss of 0.2 mg. An in vitro interaction assay with blood must also be done in future studies considering the entire system of ADV; because we know the damage to blood cells by the mechanical action of the pivot system and the degree of toxicity of the particles released into the circuit by abrasive wear.

**Acknowledgements**

The guiding and co supervisors, the Laboratory Associated Sensors and Materials (LAS) of theInstitute National for Space Research (INPE), the Foundation Adib Jatene (FAJ), the Foundation Support of São Paulo (FAPESP) and Coordination Improvement Personnel Higher Education (CAPES).

rosacldesa@gmail.com

# IOP | Biomedical Materials

rosacldesa@gmail.com